\documentclass[prb,twocolumn,aps,showpacs]{revtex4}
\usepackage{graphicx}
\def\ltsim{\vbox {\hbox{\lower .8\baselineskip \hbox{$<$}} \break
                 \hbox{\lower 0.2\baselineskip \hbox{$\sim$}} } }

\begin{document}

\title{Universal and measurable entanglement entropy in the spin-boson model}

\author{Angela Kopp$^{1}$ and Karyn Le Hur$^{2,3}$}
\affiliation{$^1$ Center for Materials Theory, Rutgers University, Piscataway, New Jersey 08854, USA}
\affiliation{$^2$ Department of Physics, Yale University, New Haven, CT 06520}
\affiliation{$^3$ D\'epartement de Physique, Universit\'e de Sherbrooke, 
Sherbrooke, Qu\'ebec, Canada J1K 2R1}
 
 \date{\today} 

\begin{abstract}
We study the entanglement between a qubit and its environment  from the spin-boson model with Ohmic dissipation. Through a mapping to the anisotropic Kondo model, we derive the
entropy of entanglement of the spin $E(\alpha,\Delta,h)$, where $\alpha$ is the dissipation strength, $\Delta$ is the tunneling amplitude between qubit states, and $h$ is the level asymmetry. For $1-\alpha \gg \Delta/\omega_c$ and $(\Delta,h) \ll \omega_c$, we show that the Kondo energy scale $T_K$ controls the entanglement between the qubit and the bosonic environment ($\omega_c$ is a high-energy cutoff). For $h\ll T_K$, the disentanglement proceeds as $(h/T_K)^2$; for $h\gg T_K$, $E$ vanishes as $(T_K/h)^{2-2\alpha}$, up to a logarithmic correction. For a given $h$, the maximum entanglement occurs 
at a value of $\alpha$ which lies in the crossover regime $h\sim T_K$. We emphasize the possibility of measuring this entanglement using charge qubits subject to electromagnetic noise.
\end{abstract}

\maketitle

The concept of quantum entropy appears in multiple contexts, from black hole physics \cite{Bombelli} to quantum information theory, where it measures the entanglement of quantum states.\cite{entropy}  Prompted by the link between entanglement and quantum criticality,\cite{Osterloh} a number of researchers have begun to study the entanglement entropy of condensed matter systems.  In this Letter, we employ the spin-boson model\cite{Blume,Leggett} to describe the entanglement between a qubit (two-level system) and an infinite collection of bosons.  With an Ohmic bosonic bath, the spin-boson model undergoes a quantum phase transition of Kosterlitz-Thouless type when $\alpha-1=\Delta/\omega_c$, where $\alpha$ is the strength of the coupling to the environment, $\Delta$ is the tunneling amplitude between the qubit states, and $\omega_c \gg \Delta$ is an ultraviolet cutoff.\cite{Cha,Bray} When the two levels of the qubit are degenerate, the entanglement between the qubit and the bosons is discontinuous at this transition.\cite{Costi,Angela}  Here we report the first rigorous analytical results for the entanglement (quantum entropy) in the strongly entangled regime $1-\alpha \gg \Delta/\omega_c$.

We exploit a mapping between the spin-boson model and the anisotropic Kondo model; our results follow from the Bethe ansatz solution of the equivalent interacting resonant level model.\cite{Pono,Markus} We show that the entropy of entanglement ($E$) of the qubit with the environment is controlled by the Kondo energy scale $T_K$, which governs the low-energy regime of the Kondo problem (a strongly correlated Fermi liquid).  We derive simple universal scaling forms for the entanglement in the limits $h \ll T_K$ and $h \gg T_K$ (Fig. 1), where $h \ll \omega_c$ is the level asymmetry between qubit states.  We also observe that, for a given $h$, $E$ is maximized at a value of $\alpha$ which lies in the crossover regime $h \sim T_K$.
\begin{figure}[ht]
\includegraphics[width=2.4in,height=1.8in]{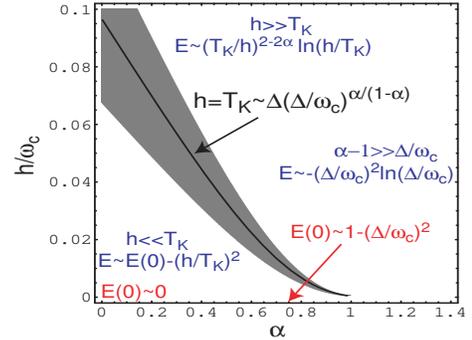}
\caption{\label{crossover} (color online) Summary of our results.  The shaded region depicts the crossover regime $h \sim T_K$.  We compute $T_K$ from the universal scaling function of Ref. \onlinecite{Young}, with $\Delta=0.1\omega_c$; for the sake of clarity we choose a large value of $\Delta$.}
\end{figure}
While the spin-boson model describes many systems of experimental interest,\cite{Leggett} the example most pertinent to this work is a noisy charge qubit, built out of Josephson junctions \cite{Schon} or metallic islands,\cite{Markus,Karyn} where the environment embodies the electromagnetic noise
stemming from Ohmic resistors in the external circuit.\cite{Rimberg}  When the qubit and the leads form a ring, the entropy of entanglement can be constructed from two measurable quantities:  the persistent current in the ring\cite{Markus,Jordan} and the charge on the dot.\cite{Lehnert} 

{\it Model and entanglement entropy.---} The Hamiltonian for the spin-boson model with a level asymmetry $h$ is:
\begin{equation}
H_{SB}=-\frac{\Delta}{2}\sigma_x+\frac{h}{2}\sigma_z+H_{osc}+\frac{1}{2}\sigma_z \sum_q \lambda_q (a_q+a_q^{\dagger}),
\label{Hsb}
\end{equation}
where $\sigma_x$ and $\sigma_z$ are Pauli matrices and $\Delta$ is the tunneling amplitude between the states with $\sigma_z=\pm 1$.  $H_{osc}$ is the Hamiltonian of an infinite number of harmonic oscillators with frequencies $\{ \omega_q \}$, which couple to the spin degree of freedom via the coupling constants $\{ \lambda_q \}$.  We assume an Ohmic heat bath with the spectral function $J(\omega) \equiv \pi \sum_q \lambda_q^2 \delta (\omega_q-\omega)=2 \pi \alpha \omega$, $\omega \ll \omega_c$.  The dimensionless parameter $\alpha$ measures the strength of the dissipation.  For $h=0$ and $\Delta/\omega_c \ll 1$, this model has a quantum critical line along the separatrix $\alpha-1=\Delta/\omega_c$.\cite{Cha} The region $\alpha-1>\Delta/\omega_c$ is a broken-symmetry phase (the ``localized'' phase) where $\Delta$ renormalizes to zero and $\lim_{h\to0}\langle \sigma_z \rangle \neq 0$; here the bosons disentangle from the spin.\cite{Angela}  The ``delocalized'' phase ($\alpha-1<\Delta/\omega_c$) is divided into two regimes by the separatrix $1-\alpha=\Delta/\omega_c$.  The localized phase can be treated by perturbation theory in $\Delta$, but in the delocalized phase this works only when $h$ is large. \cite{Markus}  We focus on the regime $1-\alpha > \Delta/\omega_c$, where the entanglement between the qubit and the environment leads to a renormalized tunneling amplitude $\Delta_{\mbox{\scriptsize ren}}<\Delta$.\cite{Cha}  
 
At zero temperature, the entanglement between two members ($A$ and $B$) of a bipartite system in the pure state $| \psi \rangle$ is given by the von Neumann entropy $E=-\mathrm{Tr} \rho_A \log_2 \rho_A=-\mathrm{Tr} \rho_B \log_2 \rho_B$, where $\rho_{A(B)}=\mathrm{Tr}_{B(A)} | \psi\rangle\langle\psi |$. \cite{entropy} If $| \psi \rangle$ is the ground state of $H_{SB}$ and $A$ is the qubit, this results in $E=-p_+ \log_2 p_+ - p_- \log_2 p_- $, where $p_{\pm}=\left( 1\pm \sqrt{\langle \sigma_x \rangle^2 + \langle \sigma_z \rangle^2}\right)/2$; $\langle \sigma_y \rangle =0$ because $H_{SB}$ is invariant under $\sigma_y \to -\sigma_y$. We present exact results for $E(\alpha,\Delta,h)$ in the regime $1-\alpha \gg \Delta/\omega_c$.  Although $E$ is defined at zero temperature, it exhibits universality that is reminiscent of thermodynamic quantities like susceptibility and specific heat. \cite{CostiZar} Recent work on the impurity entanglement in the isotropic Kondo model has emphasized universality in a similar vein. \cite{Affleck}

{\it Mapping onto the anisotropic Kondo model.---}  Our results follow from a well-known mapping between $H_{SB}$ and the anisotropic Kondo model, \cite{Anderson} defined as
\begin{eqnarray}
H_{AKM}&=&H_{\mbox{\scriptsize kin}}+\frac{J_{\perp}}{2} \sum_{kk^{\prime}} \left( c_{k\uparrow}^{\dagger}c_{k^{\prime}\downarrow} S^- + c_{k\downarrow}^{\dagger}c_{k^{\prime}\uparrow} S^+ \right) \nonumber \\
& & +\frac{J_z}{2}\sum_{kk^{\prime}} \left( c_{k\uparrow}^{\dagger}c_{k^{\prime}\uparrow}-c_{k\downarrow}^{\dagger}c_{k^{\prime}\downarrow} \right) S_z+hS_z.
\end{eqnarray}
This Hamiltonian describes the anisotropic exchange interaction between conduction electrons (labeled by the one-dimensional wave number $k$ and spin $\sigma=\uparrow,\downarrow$) and a spin-1/2 impurity.  $H_{\mbox{\scriptsize kin}}$ is the kinetic energy of the electrons.  $H_{AKM}$ and $H_{SB}$ are equivalent if we take $\Delta/\omega_c \to \rho J_{\perp}$, $\alpha \to (1+2\delta/\pi)^2$, and $h \to h$, where $\rho$ is the density of states per spin of the electrons and $\delta=\tan^{-1}(-\pi \rho J_z/4)$ is the phase shift they acquire from scattering off the impurity. \cite{Guinea2}  The region $1-\alpha>\Delta/\omega_c$ corresponds to the antiferromagnetic Kondo model, while $\alpha-1>\Delta/\omega_c$ corresponds to the ferromagnetic Kondo model. The equivalence between $H_{SB}$ and $H_{AKM}$ can be established via bosonization; a ``re-fermionization'' of $H_{SB}$ then leads to an interacting resonant level Hamiltonian \cite{Guinea2} which has been solved by Bethe ansatz.\cite{Pono} The low-energy physics of the regime $1-\alpha \gg \Delta/\omega_c$ is controlled by the Kondo scale $T_K=\Delta (\Delta/D)^{\alpha/(1-\alpha)} \sim \Delta_{\mbox{\scriptsize ren}}$, where $D$ is a high-energy cutoff. $D$ and $\omega_c$ are different but their relationship is fixed---see, {\it e.g.}, Ref. \onlinecite{Markus}.  Note that a more general expression for $T_K$ can be obtained from the renormalization group equations for $\alpha$ and $\Delta/\omega_c$; \cite{Cha,Young} in the limit $1-\alpha \to \Delta/\omega_c$, $T_K$ assumes the exponential form of the isotropic, antiferromagnetic Kondo model (Fig.~\ref{crossover}).

{\it Generalities.---}  It is clear from Eq.~(\ref{Hsb}) that $\langle \sigma_x \rangle=-2\partial E_g/\partial \Delta$ and $\langle \sigma_z \rangle=2\partial E_g/\partial h$, where $E_g$ is the ground state energy of $H_{SB}$. 
Since $H_{SB}$ and $H_{AKM}$ are related by a unitary transformation, they have the same ground state energy (up to an unimportant constant).  The field $h$ couples directly to the spin in both models, so we have $\langle S_z \rangle=\langle \sigma_z \rangle/2$.  However, a similar relationship does {\em not} hold between $\langle S_x \rangle$ and $\langle \sigma_x \rangle$, and therefore $E$ does not measure the entanglement between the Kondo impurity and the conduction band.  At $\alpha=0$, the qubit is decoupled from the environment; thus, $\langle \sigma_x \rangle=\Delta/\sqrt{h^2+\Delta^2}$ and $\langle \sigma_z \rangle=-h/\sqrt{h^2+\Delta^2}$.  With $p_+=1$ and $p_-=0$, we have $E=0$ for all values of $\Delta$ and $h$.  But when $\alpha \to 1^-$, for $h=0$ and $\Delta/\omega_c \to 0$, the system is equivalent to the antiferromagnetic SU(2) Kondo model with $J_\perp \approx J_z$ and $\langle \sigma_x \rangle \approx \langle \sigma_z \rangle = 0$, so we expect $E \to 1$, in agreement with previous Numerical Renormalization Group (NRG) results.\cite{Costi} On the other hand, we must have $E \to 0$ at large $h$, because the qubit is localized in the state with $\langle \sigma_z \rangle = -1$ and $\langle \sigma_x \rangle = 0$.  We argue that the Kondo mapping  allows us to examine how $E$ interpolates between these limits and to explore the phase diagram $(\alpha,h)$.

{\it Toulouse limit.---} First we focus on the point $\alpha=1/2$, which corresponds to the Toulouse limit of the Kondo model.\cite{Leggett,Guinea2} The resonant level is non-interacting in this limit, \cite{Guinea2} so the ground state energy is simply that of a level at energy $h$ with width $\sim T_K$.
We find
\begin{eqnarray}
\langle \sigma_z \rangle_{\alpha=1/2} & = & -\frac{2}{\pi} \tan^{-1} \left( \frac{h}{T_K} \right), \label{sigzalph1/2}\\
\langle \sigma_x \rangle_{\alpha=1/2} & = & -\frac{2}{\pi} \sqrt{\frac{T_K}{D}} \left[ 2+\ln \left( \frac{h^2+T_K^2}{D^2} \right) \right]. \label{sigxalph1/2}
\end{eqnarray}
First, consider the limit $h \ll T_K$, where $\langle \sigma_z \rangle \to -(2/\pi)(h/T_K)$ and $\langle \sigma_x \rangle \to -(4/\pi)\sqrt{T_K/D}[1+\ln (T_K/D)]$.  The result for $\langle \sigma_z \rangle$ is consistent with the Kondo ground state, where $\mathbf{S}$ is fully screened and $\langle S_z \rangle \propto h/T_K$ at small $h$.  Since both $\langle \sigma_x \rangle$ and $\langle \sigma_z \rangle$ are small, the system is close to maximal entanglement:
\begin{equation}
\hskip -0.04cm \lim_{h \ll T_K} E(1/2,\Delta,h)=E(1/2,\Delta,0)-\frac{2}{\pi^2 \ln 2} \left( \frac{h}{T_K} \right)^2,
\label{E1/2smh}
\end{equation}
where $E(1/2,\Delta,0)=1-\frac{8}{\pi^2 \ln 2}\frac{T_K}{D}\left[ 1+\ln\left(\frac{T_K}{D}\right)\right]^2$.
\begin{figure}[ht]
\includegraphics[width=2.4in,height=1.8in]{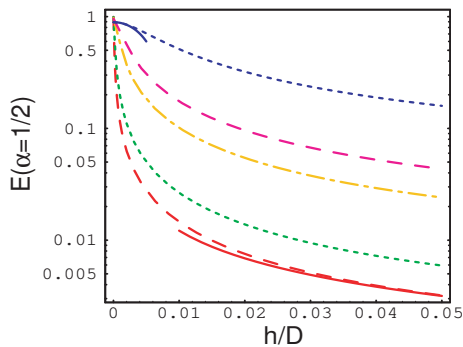}
\caption{\label{Ealpha1/2} (color online) Entropy $E(\alpha=1/2,h)$, plotted on a logarithmic scale for five values of $T_K=\Delta^2/D$; from top to bottom $T_K/D=0.005,0.001,0.0005,0.0001,0.00005$.  The solid lines show the asymptotes found in Eqs. (\ref{E1/2smh}) and (\ref{E1/2lgh}).}
\end{figure}
We have argued that $E(h=0) \to 1$ as $\alpha \to 1^-$; since the correction is already small at $\alpha=1/2$, we anticipate that $E$ varies smoothly from $\alpha=1/2$ to $\alpha=1$ at $h=0$.  The second term in Eq.~(\ref{E1/2smh}) is a universal function of $h/T_K$, with a quadratic dependence on energy that arises from the Kondo Fermi liquid behavior of $\langle \sigma_z \rangle$. In the opposite limit $h \gg T_K$, we find that $\langle \sigma_z \rangle \to -1+2T_K/(\pi h)$ and $\langle \sigma_x \rangle \to -(4/\pi) \sqrt{T_K/D} \ln (h/D)$.  Again, the leading $h$-dependence of $E$ has a universal form dictated by $\langle \sigma_z \rangle$:
\begin{equation}
\lim_{h \gg T_K}E(1/2,\Delta,h)=\frac{1}{\pi \ln 2} \left( \frac{T_K}{h} \right) \ln \left( \frac{h}{T_K} \right).
\label{E1/2lgh}
\end{equation}
Because of the logarithmic correction, $E$ approaches zero slowly at large $h$.  Plots of $E$ over the full range of $h$ are shown in Fig.~\ref{Ealpha1/2} for several values of $T_K$.  

{\it Away from $\alpha=1/2$.---}  The Bethe ansatz solution of the interacting resonant level model provides exact solutions for $\langle \sigma_z \rangle$ and $\langle \sigma_x\rangle$ in the delocalized realm $1-\alpha \gg \Delta/\omega_c$.\cite{Pono,Markus}   While the general expressions are quite complicated, we derive simple {\it  scaling} forms for the entanglement entropy in the limits $h \ll T_K$ and $h \gg T_K$.

For $h \ll T_K$ we find
\begin{equation}
\lim_{h \ll T_K} \langle \sigma_z \rangle = -\frac{2e^{\frac{b}{2(1-\alpha)}}}{\sqrt{\pi}}  \frac{\Gamma[1+1/(2-2\alpha)]}{\Gamma[1+\alpha/(2-2\alpha)]} \left( \frac{h}{T_K} \right),
\end{equation}
where $b=\alpha \ln \alpha + (1-\alpha) \ln (1-\alpha)$.  Again we have $\langle \sigma_z \rangle \propto h/T_K$ at small $h$, in keeping with the Kondo ground state.  The leading $h$-dependence of $\langle \sigma_x \rangle$ is of order $h^2$ and therefore negligible, which leaves
\begin{equation}
 \lim_{h \ll T_K} \langle \sigma_x \rangle = \frac{1}{2\alpha-1} \frac{\Delta}{\omega_c}+C_1(\alpha) \frac{T_K}{\Delta},
 \label{sigxh0}
 \end{equation}
with $C_1(\alpha)  =  \frac{e^{-b/(2-2\alpha)}}{\sqrt{\pi}(1-\alpha)} \frac{\Gamma[1-1/(2-2\alpha)]}{\Gamma[1-\alpha/(2-2\alpha)]}$.  As $\alpha \to 0$, $T_K \to \Delta$ and $C_1(0)=1$, so we recover the exact result $\langle \sigma_x \rangle_{\alpha=h=0}=1$ up to a correction of order $\Delta/\omega_c$.  This ensures $E\rightarrow 0$ for $\alpha\rightarrow 0$.  As we turn on the coupling to the environment, we introduce some uncertainty in the direction of the spin and $\langle \sigma_x \rangle$ progressively decreases. For $\alpha < 1/2$, the monotonic decrease of $T_K/\Delta$ dominates.  For $\alpha>1/2$, the first term in Eq.~(\ref{sigxh0}) dominates and we have  $\langle \sigma_x \rangle \sim \Delta/\omega_c \approx 0$.  The smallness of $\langle \sigma_x \rangle$ in this regime reflects the loss of coherent Rabi oscillations\cite{Leggett} that occurs at the dynamical crossover $\alpha=1/2$.  Note that $\langle \sigma_x \rangle$ remains analytic:  in the limit $\alpha \to 1/2$, we take $C_1(\alpha) = (4/\pi) \Gamma(1-2\alpha) \to 4/(\pi (1-2\alpha))$ and use the identity $D(\alpha=1/2)=4\omega_c/\pi$ to find $\langle \sigma_x \rangle \to -(4/\pi) \sqrt{T_K/D} \ln (T_K/D)$, in agreement with Eq.~(\ref{sigxalph1/2}). 

\begin{figure}
\includegraphics[width=1.5in]{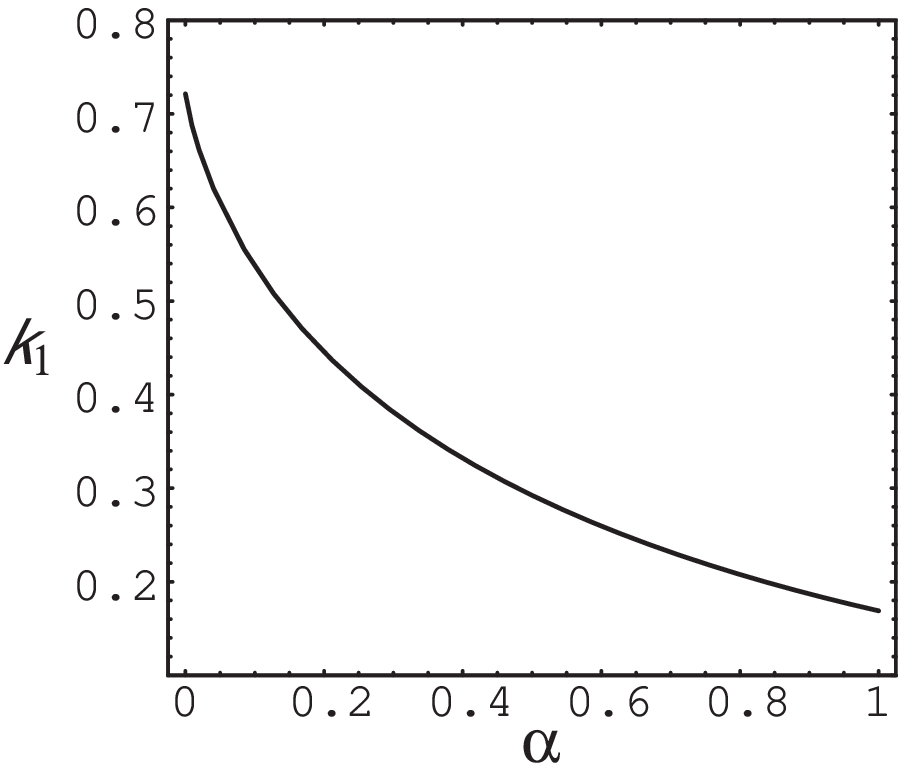} \includegraphics[width=1.5in]{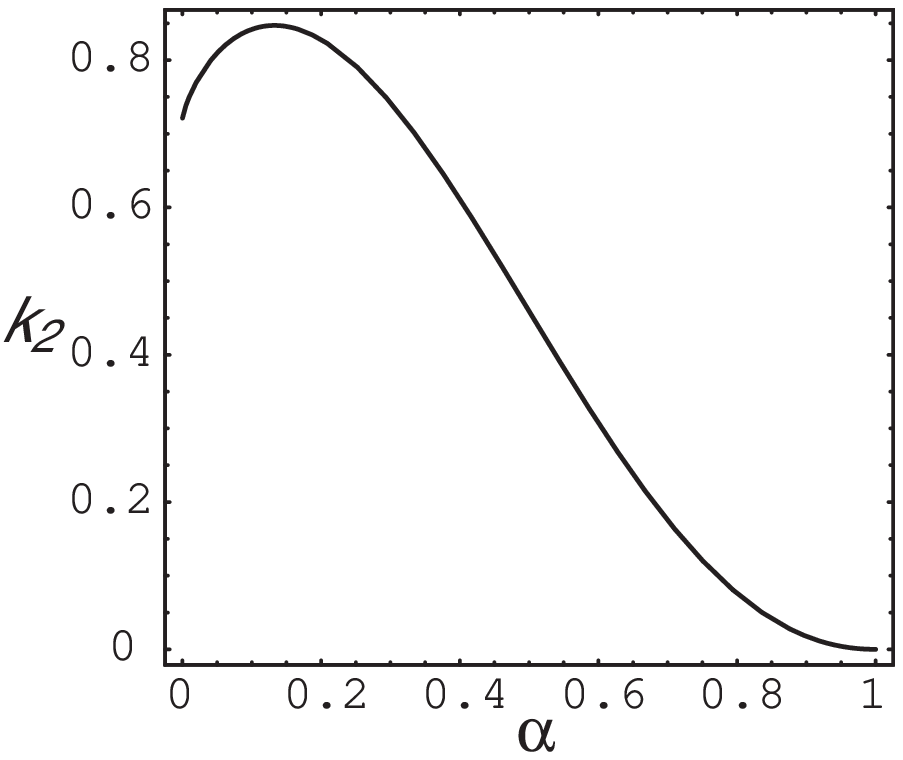}
\caption{\label{k1k2} The coefficients $k_1(\alpha)$ (left) and $k_2(\alpha)$ (right).}
\end{figure}

Now we focus on the region away from $\alpha=0$, where $T_K \ll \Delta$ and the system is strongly entangled at $h=0$.  Here we can generalize Eq.~(\ref{E1/2smh}):
\begin{equation}
\lim_{h \ll T_K \ll \Delta} E(\alpha,\Delta,h)=E(\alpha,\Delta,0)-k_1(\alpha) \left( \frac{h}{T_K} \right)^2.
\label{Esmh}
\end{equation}
The coefficient $k_1(\alpha)$ (Fig.~\ref{k1k2}) is given by
\begin{equation}
k_1(\alpha)=\frac{2 e^{\frac{b}{1-\alpha}}}{\pi \ln 2} \left( \frac{\Gamma[1+1/(2-2\alpha)]}{\Gamma[1+\alpha/(2-2\alpha)]} \right)^2,
\end{equation}
and $E(\alpha,\Delta,0)=1-\frac{1}{2 \ln 2} \left( \frac{1}{2\alpha-1} \frac{\Delta}{\omega_c}+C_1(\alpha) \frac{T_K}{\Delta} \right)^2$.
The $(h/T_K)^2$ scaling, which is a feature of the Fermi liquid fixed point, persists for $\alpha \neq 1/2$.  Note that the scaling with $h/T_K$ is determined entirely by $\langle \sigma_z \rangle$, while the 
(non-universal) contribution at $h=0$ arises from $\langle \sigma_x \rangle$.  We also observe that $E(h=0)$ saturates at maximum entropy for $\alpha>1/2$, where the leading correction is of order $(\Delta/\omega_c)^2$.  The plateau for $\alpha>1/2$ demonstrates a link between entanglement entropy and decoherence.  In the limit $\alpha \to 0$, the leading $h$-dependence is still quadratic, but with a non-universal pre-factor.

For $h \gg T_K$ we have
\begin{eqnarray}
\hskip -0.4cm \lim_{h \gg T_K} \langle \sigma_z \rangle & = & -1+\left(\frac{1-2\alpha}{2}\right) C_2(\alpha) \left( \frac{T_K}{h} \right)^{2-2\alpha} \\
\hskip -0.4cm \lim_{h \gg T_K} \langle \sigma_x \rangle & = & \frac{1}{2 \alpha -1} \frac{\Delta}{\omega_c} + C_2(\alpha) \frac{T_K}{\Delta} \left( \frac{T_K}{h} \right)^{1-2\alpha},
\end{eqnarray}
where $C_2(\alpha)=\frac{2 e^{-b}}{\sqrt{\pi} (1-2\alpha)} \frac{\Gamma(3/2-\alpha)}{\Gamma(1-\alpha)}$.
In the limit $\alpha \to 1/2$, $\langle \sigma_x \rangle$ contains two additional terms which conspire with the other terms to produce the logarithm of Eq.~(4); we do not write them explicitly because they cancel each other for $\alpha \neq 1/2$.  In the regime $T_K \ll h \ll \Delta$ we find 
\begin{equation}
\hskip -0.12cm \lim_{T_K \ll h \ll \Delta}E(\alpha,\Delta,h)=k_2(\alpha)  \ln \left( \frac{h}{T_K} \right)\left( \frac{T_K}{h} \right)^{2-2\alpha},
\label{Elgh}
\end{equation}
where the pre-factor (Fig.~\ref{k1k2}) is given by
\begin{equation}
 k_2(\alpha)=\frac{(1-\alpha)e^{-b}}{\sqrt{\pi} \ln 2} \frac{\Gamma(3/2-\alpha)}{\Gamma(1-\alpha)} \; .
\end{equation}
As in Eq.~(\ref{E1/2lgh}), the universal scaling function follows from the high-field response of $\langle \sigma_z \rangle$.

In the localized phase we obtain $\langle \sigma_z \rangle \approx -1$ and $\langle \sigma_x \rangle \approx 0$---and therefore $E \approx 0$---for infinitesimal $h$.  Since dissipation localizes the spin in the "down" state for $h=0^+$, we do not expect the entropy to depend strongly on external field; so it makes sense that $k_2 \to 0$ as $\alpha \to 1$. Deep in the localized phase, for $\alpha-1\gg \Delta/\omega_c$, we find from perturbation theory\cite{Markus} that  $E \sim -(\Delta/\omega_c)^2 \ln (\Delta/\omega_c)$ to leading order for all $h$ (Fig. 1). As we approach the critical line $\alpha-1 = \Delta/\omega_c$ from the localized side, this behavior is replaced by $E \sim -(\Delta/\omega_c) \ln (\Delta/\omega_c)$.\cite{Angela}
\begin{figure}
\includegraphics[width=2.8in,height=1.8in]{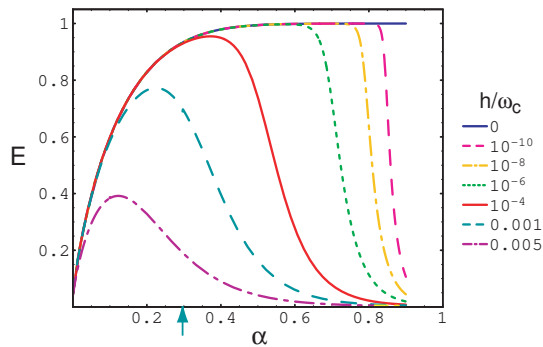}
\caption{\label{Evsalpha} (color online) $E(\alpha,\Delta=0.01 \omega_c,h)$ versus $\alpha$ at several values of $h$.  At $h=0$, we check that $E \propto \alpha$ in the limit $\alpha \to 0$. The arrow marks the value of $\alpha$ at which $T_K=0.001\omega_c$; we see that for $h=0.001 \omega_c$, $E$ is maximized near this point.}
\end{figure}

We use the full solutions\cite{Pono,Markus} for $\langle \sigma_x \rangle$ and $\langle \sigma_z \rangle$ to plot $E$ versus $\alpha$ for various $h$ in Fig.~\ref{Evsalpha}.  The entropy increases monotonically when $h=0$: it is linear in $\alpha$ near $\alpha=0$, and it saturates at $E \approx 1$ for $\alpha>1/2$, as discussed above.  As $h$ increases from zero, $E$ exhibits a maximum at progressively smaller values of $\alpha$, in agreement with previous NRG results. \cite{Costi} In our view, this maximum signifies the crossover $h \sim T_K$ (Fig.~\ref{Evsalpha}).  If $h=0$, the entropy $E$ is driven to zero by dissipation, and we observe instead a sharp non-analyticity at the phase transition.\cite{Angela}

{\it Experiments.---} An important open question in the study of quantum entanglement is whether it can be measured experimentally.  The model considered here is realized in noisy charge qubits, composed either of a large metallic dot \cite{Markus,Karyn} (the single electron box) or a superconducting island \cite{Schon} (the Cooper pair box).  The gate voltage controls the level asymmetry $h$, and $\Delta$ corresponds to the tunneling amplitude between the dot and the lead or the Josephson coupling energy of the junction. If the gate voltage source is placed in series with an external impedance, voltage fluctuations will give rise to dissipation even at zero temperature.\cite{Karyn2}  The parameter $\alpha$ can be varied {\em in situ} when a two-dimensional electron gas acts as the Ohmic dissipative environment. \cite{Rimberg} Here, $E$ depends only on $\langle \sigma_x \rangle$ and $\langle \sigma_z \rangle$, so it can be constructed from physical observables.  While these quantities would obviously be measured at finite temperature, we assume it is possible to recover the ground state density matrix by extrapolating them to their zero-temperature values.  Charge measurements \cite{Lehnert} yield the quantity $\langle \sigma_z \rangle$, which represents the occupation of the dot or island.  In a ring geometry, the application of a magnetic flux generates a persistent current that is proportional to the observable $\langle \sigma_x \rangle$.\cite{Markus}  Another promising system is the atomic quantum dot, which also permits experimental control of the coupling between the dot and the bosonic reservoir. \cite{Recati}

{\it Conclusion.---}  We have provided quantitative predictions for the entropy of entanglement
of the spin in the spin-boson model. This entropy exhibits universal behavior in the delocalized phase, governed by the Fermi liquid fixed point of the equivalent anisotropic Kondo system.  We have also described an experimental setup capable of testing our predictions; such measurements would provide an empirical proof of the existence of entanglement entropy.   Although the presence of dissipation in charge qubits makes them unlikely candidates for a functioning quantum computer, we have shown that they can be used to explore interesting links between quantum entanglement, decoherence, and quantum phase transitions.  This work might be extended to two noisy qubits.


{\it Acknowledgments.---} We acknowledge fruitful discussions with I. Affleck, M. B\" uttiker, S. Chakravarty, S. Girvin, A. Leggett, and A. Nevidomskyy.  We are also grateful to the Aspen Center for Physics where this work has been developed.

\end{document}